\documentstyle{article}

\newcommand{\eps}{\varepsilon}

\newtheorem{proposition}{Proposition}[section]
\newtheorem{theorem}[proposition]{Theorem}
\newtheorem{corollary}[proposition]{Corollary}
\newtheorem{lemma}[proposition]{Lemma}

\newcommand{\art}[6]{{\rm #1, \rm #2, \it #3 \bf #4 \rm (#5), \mbox{#6}.}}

\newcommand{\book}[3]{{\rm #1, \it #2, \rm #3.}}

\begin{document}

\centerline{\Large{\bf  SPINORS, SPIN COEFFICIENTS AND
}}

\centerline{\Large{\bf  LANCZOS POTENTIALS
}}

\

\centerline{{\bf F. Andersson} and {\bf S. B. Edgar}}

\centerline{ \it Department of Mathematics,
 Link\"oping University,}

\centerline{\it S581 83 Link\"oping,}

\centerline{\it Sweden. }

\centerline{Email: frand@mai.liu.se, bredg@mai.liu.se}

\

\begin{abstract}

It has been demonstrated, in a number of special situations, that the
spin coefficients of a canonical spinor dyad can be used to define a 
Lanczos potential of the Weyl curvature spinor. In this paper we 
explore some of these potentials and show that they can be defined 
directly from the spinor dyad in a very simple way, but that the 
results do not generalize significantly, in any direct manner. A link
to metric, asymmetric, curvature-free connections, which suggests a 
more natural relationship between the Lanczos potential and spin 
coefficients, is also considered.

\end{abstract}

\section{INTRODUCTION}

Throughout this paper we will use conventions and definitions from 
\cite{PR1}. Let $W_{ABCD}$ be an arbitrary symmetric spinor; a 
symmetric spinor $L_{ABCA^{\prime}}$ is said to be a Lanczos spinor
potential of $W_{ABCD}$ if
\begin{equation}
	W_{ABCD}=2\nabla_{(A}{}^{A^{\prime}}L_{BCD)A^{\prime}}.
	\label{WL}
\end{equation}
Illge \cite{Illge} has shown that such a Lanczos potential always exists
locally (of course this requires a four dimensional spacetime with a 
metric of Lorentz signature). The Lanczos spinor potential is 
essentially the spinor analogue of the Lanczos {\em tensor} potential
\cite{Lanczos} for which Bampi and Caviglia \cite{BC} have given an
existence proof, by tensor methods in analytic, four dimensional spacetimes
independent of metric signature. Note that no assumptions about the
differential properties of $W_{ABCD}$ are made. Illge has also shown that
the solution of (\ref{WL}) is far from unique. For a recent summary of
properties of the Lanczos potential, see \cite{EH}. Of particular interest
is the case when $W_{ABCD}=\Psi_{ABCD}$ i.e., the Weyl curvature spinor.
There exists no algorithm for finding Lanczos potentials in general
spacetimes, but in certain special situations some algorithms have been
found (see e.g., \cite{PCL}, \cite{DK}, \cite{EH} and \cite{NV}).

Let $(o^{A},\iota^{A})$ be a spinor dyad, normalized so that 
$o_{A}\iota^{A}=1$. It is then conventional to define the 8 dyad
components of the Lanczos potential as
\begin{eqnarray}
 L_{0}=L_{ABCA^{\prime}}o^{A}o^{B}o^{C}o^{A^{\prime}} & \quad L_{4}=
 L_{ABCA^{\prime}}o^{A}o^{B}o^{C}\iota^{A^{\prime}} \nonumber \\ L_{1}=
 L_{ABCA^{\prime}}o^{A}o^{B}\iota^{C}o^{A^{\prime}} & \quad L_{5}=
 L_{ABCA^{\prime}}o^{A}o^{B}\iota^{C}\iota^{A^{\prime}} \nonumber \\
 L_{2}=L_{ABCA^{\prime}}o^{A}\iota^{B}\iota^{C}o^{A^{\prime}} & \quad 
 L_{6}=L_{ABCA^{\prime}}o^{A}\iota^{B}\iota^{C}\iota^{A^{\prime}}
 \nonumber \\ L_{3}=L_{ABCA^{\prime}}\iota^{A}\iota^{B}
 \iota^{C}o^{A^{\prime}} & \quad L_{7}=L_{ABCA^{\prime}}
 \iota^{A}\iota^{B}\iota^{C}\iota^{A^{\prime}}. \label{eq:L0-L7}
\end{eqnarray}
Note that in \cite{PCL} and \cite{LMNR}, these Lanczos scalars are 
defined from the {\em tensor} version of the Lanczos potential; 
therefore the scalars used in \cite{PCL} and \cite{LMNR} are defined 
using the opposite sign compared to our spinor definition.

Now, (\ref{WL}) can be written as 5 scalar equations in NP-formalism
\cite{PCL}, \cite{LMNR}
\begin{eqnarray}
	\frac{1}{2}\Psi_{0} & = & \delta L_{0}-D 
	L_{4}-(\bar{\alpha}+3\beta -\bar{\pi})L_{0} +3\sigma 
	L_{1}+(3\eps-\bar{\eps}+\bar{\rho})L_{4}-3\kappa L_{5} \nonumber \\ 
	2\Psi_{1} & = & 3\delta L_{1}-3D L_{5}-\bar{\delta}L_{4}+\Delta 
	L_{0}-(3\gamma+\bar{\gamma}+3\mu-\bar{\mu})L_{0} \nonumber \\ & & 
	-3(\bar{\alpha}+\beta-\bar{\pi}-\tau)L_{1}+6\sigma L_{2}+(3\alpha-
	\bar{\beta}+3\pi+\bar{\tau})L_{4} \nonumber \\ & & +3(\eps-\bar{\eps}
	+\bar{\rho}-\rho)L_{5}-6\kappa L_{6} \nonumber \\Ê\Psi_{2} & = &
	\delta L_{2}-D L_{6}-\bar{\delta}L_{5}+\Delta L_{1}-\nu L_{0}-(2\mu
	-\bar{\mu}+\gamma+\bar{\gamma})L_{1} \nonumber \\ & & -(\bar{\alpha}-
	\beta-\bar{\pi}-2\tau)L_{2}+\sigma L_{3}+\lambda L_{4}+(\alpha-
	\bar{\beta}+2\pi+\bar{\tau})L_{5} \nonumber \\Ê& & -(\eps+\bar{\eps}-
	\bar{\rho}+2\rho)L_{6}-\kappa L_{7} \nonumber \\ 2\Psi_{3} & = &
	\delta L_{3}-D L_{7}-3\bar{\delta}L_{6}+3\Delta L_{2}-6\nu L_{1}
	+3(\bar{\mu}-\mu+\gamma-\bar{\gamma})L_{2} \nonumber \\ & & -
	(\bar{\alpha}-3\beta-3\tau-\bar{\pi})L_{3}+6\lambda L_{5}-3(\alpha+
	\bar{\beta}-\bar{\tau}-\pi)L_{6} \nonumber \\Ê& & -(3\eps+\bar{\eps}
	-\bar{\rho}+3\rho)L_{7} \nonumber \\ \frac{1}{2}\Psi_{4} & = &
	\Delta L_{3}-\bar{\delta}L_{7}-3\nu L_{2}+(\bar{\mu}+3\gamma-
	\bar{\gamma})L_{3}+3\lambda L_{6}-(3\alpha+\bar{\beta}-\bar{\tau})
	L_{7}
	\label{}
\end{eqnarray}
for the Lanczos scalars; these are called the NP Weyl-Lanczos equations.
\footnote{Unfortunately these equations contain some misprints in both
\cite{PCL} and \cite{LMNR}; in \cite{DK} the equations (\ref{WL}) subject
to the Lanczos differential gauge $\nabla^{AA^{\prime}}L_{ABCA^{\prime}}=0$
are quoted in NP-formalism and it should be noted that the equations occur
there with the opposite sign compared to above.}

In \cite{PCL} it has been pointed out that if the Lanczos scalars in the NP
Weyl-Lanczos equations are replaced by the spin coefficients according to the
scheme
\begin{eqnarray}
	L_{0} & = & \frac{\kappa}{2}\;,\quad\;\;\;\! L_{4}\;\;=\;\;\frac{\sigma}{2} 
	\nonumber \\ L_{1} & = & \frac{\rho}{6}\;,\quad\;\;\; L_{5}\;\;=\;\;  
	\frac{\tau}{6} \nonumber \\ L_{2} & = & 
	-\frac{\pi}{6}\;,\quad L_{6}\;\;=\;\;-\frac{\mu}{6} 
	\nonumber \\ L_{3} & = & -\frac{\lambda}{2}\;,\quad L_{7}
	\;\;=\;\;-\frac{\nu}{2}
	\label{Npot}
\end{eqnarray}
then the resulting equations for the spin coefficients can be shown to be 
satisfied in Petrov type N spaces with a suitably chosen dyad, by 
virtue of the Ricci equations of the NP-formalism. Furthermore it is 
shown that the {\em different} replacement
\begin{eqnarray}
	L_{0} & = & \kappa\;,\quad\;\;\; L_{4}\;\;=\;\;\sigma 
	\nonumber \\ L_{1} & = & \frac{\rho}{3}\;,\quad \;\;\;\!L_{5}\;\;= 
	\;\;\frac{\tau}{3} \nonumber \\ L_{2} & = & 
	-\frac{\pi}{3}\;,\quad L_{6}\;\;=\;\;-\frac{\mu}{3} 
	\nonumber \\ L_{3} & = & -\lambda\;,\quad\; L_{7}\;\;= 
	\;\;-\nu,
	\label{IIIpot}
\end{eqnarray}
again because of the Ricci equations, satisfies the Weyl-Lanczos 
equations of a type III spacetime, in a suitably chosen dyad.

These results can both be extended to Petrov type 0 spaces, but 
neither of these results --- nor any obvious modification --- is applicable
to any other spaces.

In Section 2 we show how the results in \cite{PCL} are actually 
consequences of a simple spinor ansatz
$$
 L_{ABC}{}^{A^{\prime}}=\nabla_{(A}{}^{A^{\prime}}\Bigl(g\,o_{B}
 \iota_{C)}\Bigr)
$$
where $g$ is an arbitrary function, and a very simple spinor calculation. We
show that this also makes it clear why the applicability of this ansatz is
restricted to spacetimes with very special Weyl spinors.

In \cite{LMNR} a `generalised Weyl-Lanczos equation'
\begin{equation}
		\Psi_{ABCD}=2f\nabla_{(A}{}^{A^{\prime}}L_{BCD)A^{\prime}}
	\label{genWL1}
\end{equation}
was proposed for type D vacuum spacetimes. Possible choices are
$f=\Psi_{2}^{\frac{2}{3}}$ and $f=\Psi_{2}^{\frac{1}{3}}$; for these choices
it was shown that the Lanczos scalars can be chosen as $f^{-1}$ times linear
combinations of spin coefficients, in an analogous manner as for the type
III, N and 0 potentials found in \cite{PCL}. In Section 3 it is shown how
these `generalised Lanczos potentials' can also be presented as a simple,
direct ansatz via a spinor dyad. It is then obvious that this particular
algorithm will {\em only} work for Petrov type D, vacuum spaces (or
insignificant non-vacuum generalizations).

Although the particular identifications between the Lanczos scalars 
and the spin coefficients presented in \cite{PCL} and \cite{LMNR} 
seem to be simply mathematical curiosities which are incapable of any
significant direct generalization, it is important to note that there
are fundamental structural links between Lanczos potentials and spin
coefficients, as we discuss in Section 4. There we demonstrate such a link
for a subclass of Kerr-Schild spacetimes, and moreover show the 
relationship to curvature-free asymmetric connections. More precisely, 
given a spinor dyad $(o^{A},\iota^{A})$, the spin coefficients of this 
dyad are given by the components of the spinor
$$
 \gamma_{CBAA'}=\iota_{C}\nabla_{AA'}o_{B}-o_{C}\nabla_{AA'}\iota_{B}.
$$
We prove that in a Kerr-Schild spacetime with metric $g_{ab}=\eta_{ab}
+fl_{a}l_{b}$ where $\eta_{ab}$ is a flat metric and $l^{a}$ is a geodesic,
shear-free null-vector, there exists a spinor dyad $(o^{A},\iota^{A})$
such that $L_{ABCA'}=\frac{1}{2}\gamma_{(ABC)A'}$ is a Lanczos 
potential of the Weyl spinor.

\section{LANCZOS POTENTIALS FOR TYPE III, N AND 0 WEYL SPINORS}

Of course, the first thought of how to construct an algorithm for the 
Lanczos potential would be to think in terms of derivatives of the 
metric since, in linearized theory we can obtain the Weyl tensor as a 
certain combination of the metrics second derivatives and consequently 
we can obtain a `linearized Lanczos potential' expressed in the first 
partial derivatives of the metric components in a suitable coordinate 
system. However, in the full, non-linear theory the corresponding equation
would contain other terms and so, the translation back to covariant derivatives
would be difficult, even impossible if the remainder terms failed to cancel.

Instead of the metric, one could next think in terms of derivatives 
of the tetrad vectors. However, it is clear that when working with the 
Lanczos potential (in four dimensional spacetimes) the spinor 
structure is much simpler than the tensor structure; one needs only 
to compare the simple defining equation (\ref{WL}) in spinors with the 
equivalent complicated defining equation (see \cite{Lanczos}, \cite{EH})
in tensors. Therefore we will try to build our Lanczos potential from the
spinor dyad $(o^{A},\iota^{A})$ where $o_{A}\iota^{A}=1$.

One of the simplest constructions one could think of is 
\begin{equation}
	L_{ABC}{}^{A^{\prime}}=\nabla_{(A}{}^{A^{\prime}}\Bigl(g\,o_{B}
	\iota_{C)}\Bigr)
	\label{potcand}
\end{equation}
where $g$ is an arbitrary function of the points in spacetime. Expanding
the Weyl spinor in this dyad gives
\begin{eqnarray}
 \Psi_{ABCD} & = & \Psi_{0}\iota_{A}\iota_{B}\iota_{C}\iota_{D}-4\Psi_{1}
 o_{(A}\iota_{B}\iota_{C}\iota_{D)}+6\Psi_{2}o_{(A}o_{B}\iota_{C}\iota_{D)}
 \nonumber \\ & & -4\Psi_{3}o_{(A}o_{B}o_{C}\iota_{D)}+\Psi_{4}o_{A}o_{B}
 o_{C}o_{D}. \nonumber
\end{eqnarray}
Thus, we obtain from (\ref{potcand})
\begin{eqnarray}
	2\nabla_{(A}{}^{A^{\prime}}L_{BCD)A^{\prime}} & = & -2
	\nabla_{A^{\prime}(D}\nabla_{A}{}^{A^{\prime}}\bigl(g\,o_{B}\iota_{C)}
	\bigr)=2g\Psi_{(ABC}{}^{E}\bigl(o_{D)}\iota_{E}+\iota_{D)}o_{E}\bigr)
	\nonumber \\ & = & 2g\bigl(\Psi_{0}\iota_{A}\iota_{B}\iota_{C}
	\iota_{D}-3\Psi_{1}o_{(A}\iota_{B}\iota_{C}\iota_{D)}+\Psi_{1}o_{(A}
	\iota_{B}\iota_{C}\iota_{D)} \nonumber \\ & & +3\Psi_{2}o_{(A}o_{B}
	\iota_{C}\iota_{D)}-3\Psi_{2}o_{(A}\iota_{B}\iota_{C}o_{D)}
	+3\Psi_{3}o_{(A}o_{B}\iota_{C}o_{D)} \nonumber \\ & & -\Psi_{3}o_{(A}
	o_{B}o_{C}\iota_{D)}-\Psi_{4}o_{A}o_{B}o_{C}o_{D}\bigr) \nonumber \\
	& = & 2g\Psi_{0}\iota_{A}\iota_{B}\iota_{C}\iota_{D}-4g\Psi_{1}o_{(A}
	\iota_{B}\iota_{C}\iota_{D)}+4g\Psi_{3}o_{(A}o_{B}o_{C}\iota_{D)}
	\nonumber \\ & & -2g\Psi_{4}o_{A}o_{B}o_{C}o_{D}.
	\label{}
\end{eqnarray}
From this calculation we can draw a number of conclusions:
 
\underline{If $\Psi_{ABCD}$ is type 0}, then clearly $2\nabla_{(A}
{}^{A^{\prime}}L_{BCD)A^{\prime}}=0=\Psi_{ABCD}$ for all $g$ so
$L_{ABCA^{\prime}}$, as given by (\ref{potcand}), is a Lanczos potential
for the Weyl spinor for any choice of $g$.

\underline{If $\Psi_{ABCD}$ is type N}, then we can choose $o^{A}$ as the
principal spinor of $\Psi_{ABCD}$. Then $\Psi_{0}=\Psi_{1}=\Psi_{2}=
\Psi_{3}=0$, so
$$
 2\nabla_{(A}{}^{A^{\prime}}L_{BCD)A^{\prime}}=-2g\Psi_{4}o_{A}o_{B}o_{C}o_{D}
 =-2g\Psi_{ABCD}.
$$
This means that $L_{ABCA^{\prime}}$ is a Lanczos potential for the Weyl
spinor if and only if $g=-\frac{1}{2}$. Note that this holds for any choice
of $\iota^{A}$, as long as $o_{A}\iota^{A}=1$. For this choice,
$L_{ABCA^{\prime}}$ can easily be seen to coincide with the potential
(\ref{Npot}) originally found in \cite{PCL}.

\underline{If $\Psi_{ABCD}$ is type III}, then we can choose $o^{A}$ as
the repeated principal spinor of $\Psi_{ABCD}$ and $\iota^{A}$ as the 
other principal spinor so that $\Psi_{0}=\Psi_{1}=\Psi_{2}=\Psi_{4}=0$.
Hence,
$$
 2\nabla_{(A}{}^{A^{\prime}}L_{BCD)A^{\prime}}=4g\Psi_{3}o_{(A}o_{B}
 o_{C}\iota_{D)}=-g\Psi_{ABCD}
$$
so $L_{ABCA^{\prime}}$ as given by (\ref{potcand}) is a Lanczos potential for
the Weyl spinor if and only if $g=-1$. This choice is easily seen to coincide
with (\ref{IIIpot}) originally found in \cite{PCL}.

\underline{If $\Psi_{ABCD}$ is type D}, then we can choose $o^{A}$ and 
$\iota^{A}$ as the repeated principal spinors of $\Psi_{ABCD}$ 
so that $\Psi_{0}=\Psi_{1}=\Psi_{3}=\Psi_{4}=0$. Thus, in this case
$$
 2\nabla_{(A}{}^{A^{\prime}}L_{BCD)A^{\prime}}=0
$$
for all functions $g$. Therefore we cannot use $L_{ABCA^{\prime}}$ given
by (\ref{potcand}) as a Lanczos potential for the Weyl spinor in type D.
We can however use it as a gauge transformation i.e., if we know a Lanczos
potential for a type D Weyl spinor, then we can add $L_{ABCA^{\prime}}$ 
given by (\ref{potcand}) to it, and the sum will still be a Lanczos potential
of the Weyl spinor.

For other, \underline{more general spacetimes}, it is easily seen 
that $L_{ABCA^{\prime}}$ as given by (\ref{potcand}) will not be a 
Lanczos potential of $\Psi_{ABCD}$ because the crucial $\Psi_{2}$-component 
vanishes for any choice of $g$.

\section{A `GENERALIZED LANCZOS POTENTIAL' FOR WEYL SPINORS OF EMPTY, TYPE
D SPACETIMES}

In \cite{LMNR} a `generalized Weyl-Lanczos equation'
\begin{equation}
		\Psi_{ABCD}=2f\nabla_{(A}{}^{A^{\prime}}L_{BCD)A^{\prime}}
	\label{genWL}
\end{equation}
is proposed for a type D vacuum spacetime. Some particular 
solutions for $L_{ABCA^{\prime}}$ are found for the choices $f=
\Psi_{2}^{\frac{2}{3}}$ and $f=\Psi_{2}^{\frac{1}{3}}$. These
`generalized Lanczos potentials' can be found after some extensive NP
calculations using both the Ricci equations and the Bianchi equations.

For $f=\Psi_{2}^{\frac{1}{3}}$ it is shown that a solution is given by
\begin{eqnarray}
	L_{i} & = & 0\;,\;\;i=0,3,4,7 \nonumber \\ L_{1} & = &
	-\frac{\eps}{3}f^{-1}\;,\;\;L_{5}\;\;=\quad -\frac{\beta}{3}f^{-1}
	\nonumber \\ L_{2} & = & -\frac{1}{3}(\pi+\alpha)f^{-1}\;,\quad
	L_{6}\;\;=\;\;-\frac{1}{3}(\mu+\gamma)f^{-1}.
	\label{}
\end{eqnarray}
However it is easy to see that these expressions can be obtained in a 
similar way as in Section 2 i.e., $L_{ABCA^{\prime}}$ is given simply 
by
\begin{equation}
	L_{ABCA^{\prime}}=-f^{-1}o_{(A}\nabla_{B|A^{\prime}|}\iota_{C)}.
	\label{}
\end{equation}
Here $o^{A}$ and $\iota^{A}$ are the principal spinors of $\Psi_{ABCD}$
(normalized so that $o_{A}\iota^{A}=1$). Obviously an alternative is
\begin{equation}
	L_{ABCA^{\prime}}=f^{-1}\iota_{(A}\nabla_{B|A^{\prime}|}o_{C)}.
	\label{}
\end{equation}
For the case $f=\Psi_{2}^{\frac{2}{3}}$ we can obtain the form for
$L_{ABCA^{\prime}}$ given in \cite{LMNR} from
\begin{equation}
	L_{ABCA^{\prime}}=3f^{-1}\Bigl(\iota_{(A}\nabla_{B|A^{\prime}|}o_{C)}
	-\iota^{D}o_{(A}\iota_{B}\nabla_{C)A^{\prime}}o_{D}\Bigr)
	\label{}
\end{equation}
or alternatively
\begin{equation}
	L_{ABCA^{\prime}}=-3f^{-1}\Bigl(o_{(A}\nabla_{B|A^{\prime}|}\iota_{C)}
	+o^{D}o_{(A}\iota_{B}\nabla_{C)A^{\prime}}\iota_{D}\Bigr)
	\label{}
\end{equation}
By calculations similar to those in the previous section it can be 
shown that these Lanczos spinors, with the appropriate choice of $f$,
satisfy (\ref{genWL}).

\section{THE SPIN-COEFFICIENTS AS LANCZOS SCALARS}

In \cite{PCL} it has been conjectured that in general the dyad 
components of a Lanczos potential will be given by Lanczos scalars that
are linear combinations of the spin-coefficients. In support of this,
the authors of \cite{PCL} have found other different identifications
\cite{ALOM} between Lanczos scalars and the spin coefficients which
`work' in a similar way for certain other very specialised spaces e.g.,
for the Schwarzschild metric in a suitably chosen dyad
$$
 L_{i}=0,\; i \neq 1,6 \;,\;\;\;L_{1}=L_{6}=\frac{2}{3}\eps
$$
It seems to us that these very special results in 
\cite{PCL} and \cite{ALOM} are not part of a larger picture, but are
simply mathematical coincidences in very special spaces where there
is comparatively little structure. We note that the result for type 
N is independent of our choice for $\iota^{A}$ whereas in type III
we have to choose $\iota^{A}$ as our second principal spinor of 
$\Psi_{ABCD}$. In both of these cases only properly weighted spin 
coefficients are used in the identification with the Lanczos scalars 
i.e., the results still have spin-boost freedom, but in some of the 
other spaces in \cite{ALOM}, such as Schwarzschild, the Lanczos 
scalars are identified with non-weighted spin coefficients, so that 
there is no remaining dyad freedom. Most crucially, a `relationship' 
between Lanczos scalars and spin coefficients which needs to change 
depending on Petrov type, and even subtypes, appears to be more of a 
curiosity than a manifestation of some genuine deep mathematical 
structure.

However, we believe that there are structural relationships still to 
be fully understood and exploited between the Lanczos potential and 
spin coefficients as we will explain below.


Noting that the successful choices for $L_{ABCA^{\prime}}$ in Section 
2, can be written
$$
 L_{ABCA^{\prime}}=c\Bigl(o_{(A}\nabla_{B|A^{\prime}|}\iota_{C)}+\iota_{(A}
 \nabla_{B|A^{\prime}|}o_{C)}\Bigr)
$$
where the constant $c$ varies with Petrov type, therefore suggests that
we try some variations e.g.,
\begin{equation}
	L_{ABCA^{\prime}}=\frac{1}{2}\Bigl(\iota_{(A}\nabla_{B|A^{\prime}|}
	o_{C)}-o_{(A}\nabla_{B|A^{\prime}|}\iota_{C)}\Bigr).
	\label{potcand2}
\end{equation}
For future convenience we put
$$
 \gamma_{ABCA^{\prime}}=\iota_{A}\nabla_{CA^{\prime}}o_{B}-o_{A}
 \nabla_{CA^{\prime}}\iota_{B}
$$
so that $L_{ABCA^{\prime}}=\frac{1}{2}\gamma_{(ABC)A^{\prime}}$. By
calculations similar to those in Section 2 we easily obtain
\begin{equation}
	2\nabla_{(A}{}^{A^{\prime}}L_{BCD)A^{\prime}}=\nabla_{(A}{}^{A^{\prime}}
	\gamma_{BCD)A^{\prime}}=\Psi_{ABCD}-2\nabla_{(A}{}^{A^{\prime}}
	o_{D}\nabla_{B|A^{\prime}|}\iota_{C)}.
	\label{}
\end{equation}
By noting the relationship $\nabla_{(A}{}^{A^{\prime}}o_{D}
\nabla_{B|A^{\prime}|}\iota_{C)}=\frac{1}{2}\gamma_{E(DA}{}^{A^{\prime}}
\gamma^{E}{}_{CB)A^{\prime}}$ it follows that 
\begin{equation}
	2\nabla_{(A}{}^{A^{\prime}}L_{BCD)A^{\prime}}=2\nabla_{(A}{}^{A^{\prime}}
	\gamma_{BCD)A^{\prime}}=\Psi_{ABCD}-\gamma_{E(DA}{}^{A^{\prime}}
	\gamma^{E}{}_{CB)A^{\prime}}.
	\label{coeffpot}
\end{equation}
Clearly, this choice of $L_{ABCA^{\prime}}$ will be a Lanczos 
potential if and only if
$$
 \gamma_{E(DA}{}^{A^{\prime}}\gamma^{E}{}_{CB)A^{\prime}}=0.
$$
The spinor $\gamma_{ABCA^{\prime}}$ is of course a familiar structure; the
dyad components $\gamma_{\alpha\beta\gamma\alpha^{\prime}}$ (Greek letters
denote dyad indices and range from 0 to 1) of $\gamma_{ABCA^{\prime}}$
yield the familiar NP spin coefficients.\footnote{Note that in \cite{PR1}
the indices of $\gamma_{\alpha\beta\gamma\alpha^{\prime}}$ are arranged
somewhat differently.} Thus, we have the following result:

\begin{lemma}
 A Lanczos potential $L_{ABCA^{\prime}}$ of the Weyl spinor can be directly
 equated to the spin coefficients $\gamma_{ABCA^{\prime}}$ i.e.,
 $L_{ABCA^{\prime}}=\frac{1}{2}\gamma_{ABCA^{\prime}}$ if and only if
 \begin{equation}
  \gamma_{E(DA}{}^{A^{\prime}}\gamma^{E}{}_{CB)A^{\prime}}=0.
  \label{Lpotcond}
 \end{equation}
\end{lemma}

The link in differential structure between equations (\ref{WL}) and 
(\ref{coeffpot}) has been commented on also by Bonanos \cite{Bonanos}.

We can write the condition (\ref{Lpotcond}) out in full in the 
familiar spin coefficient notation as
\begin{eqnarray}
	0 & = & \sigma \eps-\kappa \beta \nonumber \\ 
	0 & = & -\kappa(\mu+\gamma)+\sigma(\pi+\alpha)-\rho \beta+\tau \eps
	        \nonumber \\
	0 & = & -\kappa \nu+\sigma \lambda-\rho \mu+\tau \pi-\rho \gamma+\tau
	        \alpha-\mu \eps+\pi \beta \nonumber \\
	0 & = & -\nu(\rho+\eps)+\lambda(\tau+\beta)-\mu\alpha+\pi \gamma
	        \nonumber \\
	0 & = & -\nu\alpha+\lambda \gamma
	\label{condcomp}
\end{eqnarray}
It is clear, from consideration of familiar simple spacetimes in 
NP-formalism, that (\ref{condcomp}) does not usually hold in the familiar 
choices of spinor dyad e.g., vacuum type D with $\kappa=\sigma=\nu=
\lambda=0$ or N with $\kappa=\sigma=\tau=\pi=\lambda=0$.

However, this does not mean that Lanczos potentials cannot be 
constructed from spin coefficients in this manner; rather it poses the 
question as to whether dyads can be found in which the spin 
coefficients satisfy (\ref{condcomp}).

Recently Lanczos potentials have been found for a 
class of spacetimes --- Kerr-Schild spacetimes \cite{KSMH} in which the 
null-vector occurring in the metric is geodesic and shear-free. 
Further these Lanczos potentials have been found in the context of 
curvature-free, asymmetric metric connections. Recall that any 
metric connection $\hat{\nabla}_{AA^{\prime}}$ can be written
\begin{equation}
	\hat{\nabla}_{AA^{\prime}}\xi^{B}=\nabla_{AA^{\prime}}\xi^{B}+
	2\Gamma_{C}{}^{B}{}_{AA^{\prime}}\xi^{C}
	\label{conn}
\end{equation}
where $\Gamma_{CBAA^{\prime}}=\Gamma_{(CB)AA^{\prime}}$. Further, 
recall that a Kerr-Schild spacetime is a spacetime in which the 
metric $g_{ab}$ can be written $g_{ab}=\eta_{ab}+fl_{a}l_{b}$ for some 
function $f$, where $\eta_{ab}$ is a flat metric and $l^{a}$ is a
null-vector with respect to $g_{ab}$. In \cite{AE} the following 
theorem is proved:
\begin{theorem}
 In a Kerr-Schild spacetime the spinor
 $$
  \Gamma_{ABCA^{\prime}}=\frac{1}{2}\nabla_{(A}{}^{B^{\prime}}
  \Bigl(f\xi_{B)}\xi_{C}\xi_{A^{\prime}}\xi_{B^{\prime}}\Bigr)
 $$
 where $l_{a}=\xi_{A}\xi_{A^{\prime}}$, defines an asymmetric connection 
 $\hat{\nabla}_{AA^{\prime}}$ according to (\ref{conn}), with vanishing 
 curvature tensor. Furthermore, if the null-vector $l^{a}$ is geodesic
 and shear-free, the spinor $L_{ABCA^{\prime}}=\Gamma_{(ABC)A^{\prime}}$
 is a Lanczos potential of the Weyl spinor.
\end{theorem}
We remark that the first part of the theorem was proved by Harnett
\cite{Harnett} while Bergqvist \cite{Bergqvist} has proved the complete
theorem in the special case of the Kerr spacetime.

It is well-known that since $\hat{\nabla}_{AA^{\prime}}$ has zero 
curvature, there exists a normalized spinor dyad $(o^{A},\iota^{A})$
such that $\hat{\nabla}_{AA^{\prime}}o_{B}=\hat{\nabla}_{AA^{\prime}}
\iota_{B}=0$. From this we easily obtain the relation
$$
 \Gamma_{CBAA^{\prime}}=\frac{1}{2}(\iota_{C}\nabla_{AA^{\prime}}
 o_{B}-o_{C}\nabla_{AA^{\prime}}\iota_{B})=\frac{1}{2}
 \gamma_{CBAA^{\prime}}
$$
where $\gamma_{CBAA^{\prime}}$ are the spin coefficients of the dyad
$(o^{A},\iota^{A})$ as above. This proves the following corollary:

\begin{corollary}
 In a Kerr-Schild spacetime in which $l^{a}$ is geodesic and 
 shear-free there exists a normalized spinor dyad $(o^{A},\iota^{A})$ 
 with spin coefficients $\gamma_{ABCA^{\prime}}$, such that
 $L_{ABCA^{\prime}}=\frac{1}{2}\gamma_{(ABC)A^{\prime}}$, 
 is a Lanczos potential of the Weyl spinor.
\end{corollary}

However, it is emphasised again that neither of the elements of the 
spinor dyad of the above theorem need coincide with the principal spinors 
of the Weyl spinor or the spinor $\xi_{A}$ occurring in the metric.

Whether dyads with the required properties exist for other 
spacetimes, or indeed for all spacetimes, is an open question which we 
are at presently investigating. It seems to be in this context of 
curvature-free asymmetric connections that investigating the links 
between Lanczos potentials and spin coefficients will be most useful.

\end{document}